%Paper: hep-th/9310114
%From: SHORE@crnvma.cern.ch
%Date: Tue, 19 Oct 93 12:15:20 SET
%Date (revised): Mon, 25 Oct 93 19:06:38 SET

\null

%*******************************************************************
%FONTS

\magnification 1200

%*******************************************************************
%FOOTNOTES - insert command \eightpoint to reduce

\newskip\ttglue

\font\eightrm=cmr8
\font\eighti=cmmi8
\font\eightsy=cmsy8
\font\eightbf=cmbx8
\font\eighttt=cmtt8
\font\eightsl=cmsl8
\font\eightit=cmti8
\font\sixrm=cmr6
\font\sixbf=cmbx6
\font\sixi=cmmi6
\font\sixsy=cmsy6

\def \eightpoint{\def\rm{\fam0\eightrm}% switch to 8-point type
\textfont0=\eightrm \scriptfont0=\sixrm \scriptscriptfont0=\fiverm
\textfont1=\eighti \scriptfont1=\sixi   \scriptscriptfont1=\fivei
\textfont2=\eightsy \scriptfont2=\sixsy   \scriptscriptfont2=\fivesy
\textfont3=\tenex \scriptfont3=\tenex   \scriptscriptfont3=\tenex
\textfont\itfam=\eightit  \def\it{\fam\itfam\eightit}%
\textfont\slfam=\eightsl  \def\sl{\fam\slfam\eightsl}%
\textfont\ttfam=\eighttt  \def\tt{\fam\ttfam\eighttt}%
\textfont\bffam=\eightbf  \scriptfont\bffam=\sixbf
 \scriptscriptfont\bffam=\fivebf  \def\bf{\fam\bffam\eightbf}%
\tt \ttglue=.5em plus.25em minus.15em
\setbox\strutbox=\hbox{\vrule height7pt depth2pt width0pt}%
\normalbaselineskip=9pt
\let\sc=\sixrm  \let\big=\eightbig  \normalbaselines\rm
}
%*******************************************************************
%GREEK LETTERS
\def\a{\alpha}
\def\b{\beta}
\def\c{\gamma}
\def\d{\delta}
\def\e{\epsilon}
\def\h{\eta}

\def\l{\lambda}
\def\m{\mu}
\def\n{\nu}
\def\o{\theta}
\def\p{\pi}

\def\s{\sigma}
\def\t{\tau}

\def\w{\omega}

\def\C{\Gamma}

%*****************************************************************
%OTHER MACROS

\def\dag{\dagger}

\def\BOB{     (\ldots     )}
\def\BSB{     [\ldots     ]}
\def\BCB{    \{\ldots    \}}

%*****************************************************************
%TITLE PAGE

{\nopagenumbers

 \line{\hfil SWAT 93/9    }
 \line{\hfil UGVA-DPT 1993/08-830 }
\vskip2cm
\centerline{{\bf ``FASTER THAN LIGHT'' PHOTONS AND CHARGED BLACK HOLES}}
\vskip1.5cm
\centerline{\bf R.D. Daniels{$^*$} and G.M. Shore{$^{*\dag}$} }
\vskip0.8cm
\centerline{$^*$ {\it Department of Physics}}
\centerline{\it University College of Swansea}
\centerline{\it Singleton Park}
\centerline{\it Swansea, SA2 8PP, U.K. }
\vskip0.8cm
\centerline{$^\dag$ {\it D\'epartement de Physique Th\'eorique}}
\centerline{\it Universit\'e de Gen\`eve}
\centerline{\it 24 quai E.-Ansermet}
\centerline{\it CH1211 Gen\`eve 4, Switzerland}
\vskip1.5cm
\noindent{\bf Abstract}
\vskip0.5cm
\noindent Photons propagating in curved spacetime may, depending
on their direction and polarisation, have velocities exceeding
the ``speed of light'' $c$. This phenomenon arises through
vacuum polarisation in QED and is a tidal gravitational effect
depending on the local curvature. It implies that the Principle
of Equivalence does not hold for interacting quantum field theories
in curved spacetime and reflects a quantum violation of local
Lorentz invariance. These results are illustrated for the propagation
of photons in the Reissner-Nordstr\"om spacetime characterising a
charged black hole. A general analysis of electromagnetic as well
as gravitational birefringence is presented.

\vskip2cm
\line{SWAT 93/9       \hfil}
\line{UGVA-DPT 1993/08-830  \hfil}
\line{August 1993    \hfil}

\vfill\eject }

\pageno = 1

\noindent{\bf 1.~Introduction}
\vskip0.5cm
Photons propagating in certain curved spacetime backgrounds
may, depending on their direction and polarisation, have
velocities exceeding the ``speed of light'' $c$.
This is a quantum phenomenon induced by vacuum polarisation in
quantum electrodynamics (QED). It is a tidal gravitational effect
depending explicitly on the local curvature. The Principle of
Equivalence does not hold for QED in curved spacetime.

These remarkable results were discovered by Drummond and Hathrell
in 1980 [1]. They considered the effect of one-loop vacuum polarisation
on photon propagation in de Sitter, Schwarzschild, Robertson-Walker
and gravitational wave spacetimes. In each case except de Sitter,
for which the curvature is totally isotropic, they identified
directions and polarisations for which the photon velocity was
faster than $c$.

In this paper, we extend these results to an arbitrary (but
slowly varying) background of electromagnetic and gravitational fields,
which enables us to discuss the Reissner-Nordstr\"om spacetime.
Our results therefore describe photon propagation around a charged
black hole. We find three classes of contributions to the photon
velocity:~
(a)~the basic gravitational effect identical to the
Schwarzschild case,~
(b)~the indirect effect of the charge through its
modification of the gravitational field, and~
(c)~the contribution of the electromagnetic field itself.~
The sign of the electromagnetic contribution is always such
as to reduce the photon velocity, while the gravitational effect may
increase the velocity for certain directions and polarisations.
The most striking case is orbital (i.e.~$\phi$) directed photon
propagation, where the velocities of the radial and tangential
(i.e.~$\theta$) polarisation states are
$$\eqalignno{
v_r ~&=~ 1 ~+~ {1\over240}{\a\over\p}{1\over{(Mm)^2}}
\Bigl({2M\over r}\Bigr)^3 \biggl[1~+~ {5\over12}\Bigl(
{\a\over m^2}\Bigr)^{-1}
\Bigl({Q\over Q_0}\Bigr)^2 ~-~ {2\over3}\Bigl({Q\over Q_0}\Bigr)^2
{2M\over r} \biggr] \cr
v_{\o}  ~&=~ 1 ~+~ {1\over240}{\a\over\p}{1\over{(Mm)^2}}
\Bigl({2M\over r}\Bigr)^3 \biggl[-1~+~{13\over12}\Bigl(
{\a\over m^2}\Bigr)^{-1}
\Bigl({Q\over Q_0}\Bigr)^2 ~-~ {7\over6}\Bigl({Q\over Q_0}\Bigr)^2
{2M\over r} \biggr]
&(1.1) \cr }
$$
respectively. For radial photon propagation, the velocity is always $c$.
Here, $Q$ and $M$ are the charge and mass of the black hole, $m$ is
the electron mass and $\a$ is the fine structure constant.
We have parametrised the result in terms of the accretion limit
charge\footnote{$^*$}{\eightpoint
\noindent We use dimensionless units where
$G = \hbar = c = \e_0 = 1$ throughout.} $Q_0 = Mm/\a^{1\over2}$.
For black holes with charge $Q\simeq Q_0$, the indirect type (b)
gravitational effect of the charge (the second terms in eq.(1.1) )
is suppressed by an inverse power of the (squared) electron charge
to mass ratio, where $\a/m^2 \simeq 10^{42}$. The direct
gravitational and electromagnetic effects are then of the same order,
the relative size depending on the radial distance $r$ from the
centre of the black hole. Notice that for $Q \leq Q_0$, propagation
at the horizon $r \simeq 2M$ in the orbital direction is
superluminal.

The physical origin of this phenomenon is relatively clear[1].
In picturesque terms, vacuum polarisation in QED allows the photon
to exist as a virtual $e^+ e^-$ pair, thereby acquiring an
effective size of $O(\l_c)$, where $\l_c = m^{-1}$ is the
Compton wavelength of the electron. In an anisotropic gravitational
field\footnote{$^*$}{\eightpoint
\noindent In this paper, as in ref.[1], we are only able to make
explicit calculations in the case of slowly varying background
fields. It is therefore the anisotropy rather than the inhomogeneity
of the gravitational (or electromagnetic) field which is responsible
for the effects found here. For rapidly varying, inhomogeneous
fields similar corrections are expected, depending on derivatives
of the curvature.}, the photon is therefore sensitive to the
curvature and its characteristics of propagation may differ from those
in flat space. The polarisation dependence (gravitational
birefringence[1]) reflects the anisotropy of the background field.

Since superluminal propagation is clearly a delicate issue, we
should be perfectly clear what is being described.
We find that in a local inertial frame at a given point in
spacetime the photon momentum $k^a$ satisfies the light-cone
condition
$$
\bigl( \h_{ab} ~+~ \a \s_{ab} \bigr)~k^a k^b ~=~0
,\eqno(1.2)
$$
where $\h_{ab}$ is the Minkowski metric and $\a \s_{ab}$ is the
one-loop vacuum polarisation correction, which depends on the
Riemann curvature tensor at the given point. In the Drummond-Hathrell
calculation[1], $\s_{ab}$ is of $O\bigl((\l_c/L)^2\bigr)$,
where $L$ is the curvature scale. In familiar quantum field theoretic
terms, the pole in the photon propagator in the local Lorentz frame is
shifted from $k^2=0$ to $k^2 + \a \s_{ab}k^a k^b = 0$.

The condition (1.2) is {\it not} Lorentz invariant, since it depends
explicitly on the local curvature. As noted in ref.[1], this is the
key observation which permits propagation according to eq.(1.2)
without leading to the classic paradoxes associated with superluminal
propagation in special relativity. Patching together local Lorentz
frames to cover the manifold, we find, in contrast to the
Principle of Equivalence, that photons do not in general follow
spacetime geodesics (even to the extent to which it is possible to
describe particle trajectories in quantum theory).

We shall discuss the issues and potential paradoxes raised by these
calculations in a little more detail in sect.~4. For the moment, we
simply emphasise the generality of the phenomenon. Any interacting
quantum field theory (presumably with the exception of
conformal field theories) necessarily involves a scale which will
enter the propagators through vacuum polarisation. In turn, this
means that the theory is sensitive at the quantum level to tidal
gravitational effects depending on the local curvature. Local
Lorentz invariance is therefore lost at the quantum level.
In formal terms, this can presumably be viewed as an anomaly
in local Lorentz invariance. The Principle of Equivalence, which
depends for its usual formulation on the existence of point particles,
is therefore inapplicable to the motion of real elementary
particles in quantum field theory\footnote{$^{**}$}{\eightpoint
\noindent Similar observations on the limits of the equivalence
principle in string theory have been made recently by Mende[2].
As we see here, however, it is not necessary to go beyond quantum
field theory to encounter the limits of the point particle concept.}.
Its content reduces simply to the statement that the curved spacetime
admits everywhere a local Minkowski tangent space.

The paper is arranged as follows. In sect.~2, we describe the
effective action for QED in an arbitrary background gravitational
and electromagnetic field and, following the techniques of ref.[1],
derive the equation of motion governing photon propagation.
Special emphasis is placed on identifying the expansion parameters
and approximations used since, as we discuss in sect.~4, fundamental
questions of principle concerning the observability of this
phenomenon depend critically on the parameter range of validity
of the calculation.

In sect.~3, we consider the special case of the Reissner-Nordstr\"om
spacetime, which describes (the exterior of) a charged black hole,
and derive the light-cone condition (1.2) appropriate to different
directions and polarisations of photon propagation. The essential
results are described above.

The phenomenon of electromagnetic birefringence (polarisation
dependence of the photon velocity in a background electromagnetic
field) has been demonstrated by Adler[3] for the case of a constant
magnetic field. In that case, the photon velocities corresponding
to both transverse polarisations are found to be less than $c$.
In appendix A, we show using our formalism that this result remains
true for an arbitrary anisotropic electromagnetic background
(although we are still forced to assume only weak inhomogeneity).
Superluminal propagation is a special feature of gravity.

Finally, in sect.~4, we describe briefly why the causal paradoxes
normally associated with faster than light propagation in special
relativity are avoided by the tidal nature of the modification
(1.2) of the light cone. We discuss the question of whether
this superluminal propagation is in principle observable
and assess the possibilities of answering some of the
conceptual questions surrounding this effect by extending the
derivation to short wavelength photons with $\l \ll \l_c$.

Following the original paper of Drummond and Hathrell[1]
there have been surprisingly few papers in the literature developing
this topic. The essential results of ref.[1] were rapidly
generalised to neutrinos in a Friedmann metric by Ohkuwa[4].
Some questions were raised concerning the validity
the superluminal propagation and the compatibility with a
dispersion relation (see sect.~4) by Dolgov and Khriplovich[5].
An analogous effect, superluminal propagation in a flat spacetime
with boundaries, has been investigated by Scharnhorst[6] and
Barton[7,8] who encounter the same conceptual and interpretational
questions raised by the curved spacetime effect.
Quantum field theory, in particular the Hawking effect, in
Reissner-Nordstr\"om spacetime has been investigated independently
of the question of photon propagation by Gibbons[9].

\vfill\eject

\noindent{\bf 2.~Effective Action and Photon Propagation}
\vskip0.5cm
The description of photon propagation given here is based on the
one-loop effective action for QED in a background gravitational
and electromagnetic field. This can be calculated using, e.g.,
heat kernel techniques as an expansion in powers of the
background field curvatures, i.e.~the Riemann tensor
$R_{\m\n\s\t}$ and the electromagnetic field strength $F_{\m\n}$.
Since the derivation is well-documented elsewhere ([1] and references
therein), we simply quote the first few terms in the expansion which are
relevant to photon propagation, viz.
$$\eqalignno{
\C~~=~~&- {1\over4} \int dx \sqrt{-g}~ F_{\m\n}F^{\m\n} \cr
&+{1\over m^2} \int dx \sqrt{-g}~\Bigl[~a R F_{\m\n}F^{\m\n} +
b R_{\m\n}F^{\m\s}F^\n{}_\d + c R_{\m\n\s\t}F^{\m\n}F^{\s\t}  \cr
&\hskip3cm + d D_\m F^{\m\n} D_\s F^\s{}_\n~\Bigr] \cr
&+ {1\over m^4}\int dx \sqrt{-g}~\Bigl[~z \bigl(F_{\m\n}F^{\m\n}\bigr)^2
+ y F_{\m\n}F_{\s\t}F^{\m\s}F^{\n\t}~\Bigr]  \cr
&+~~\ldots
&(2.1) \cr }
$$
where
$$\eqalignno{
&a = -{5\over 720} {\a\over \p}  \hskip2cm
 b = {26\over 720} {\a\over \p}  \hskip2cm
 c = -{2\over 720} {\a\over \p}  \cr
&d = -{1\over 30} {\a\over \p}  \hskip2cm
 z = -{5\over 180} \a^2  \hskip1.8cm
 y = {14\over 180} \a^2
&(2.2) \cr }
$$
Of course there are further terms of $O(R^2)$ and $O(R^3)$ independent
of the electromagnetic field strength but these do not play any
r\^ole in our analysis.

Eq.(2.1) should be viewed as giving the one-loop, i.e. $O(\a)$,
quantum corrections to the quadratic $F^2$ action as an expansion
both in powers of $R/m^2$ and $\a F^2/m^4$. (Here, `$R$' and `$F$'
denote generic curvature and field strength terms.) For a
Reissner-Nordstr\"om black hole of mass $M$ and charge $Q$, these
expansion parameters are ${1\over (Mm)^2} \Bigl({2M\over r}\Bigr)^3$
and ${1\over (Mm)^2}\Bigl({Q\over Q_0}\Bigr)^2
\Bigl({2M\over r}\Bigr)^4$ respectively (cf.~eq.(1.1)).
So, at the horizon
and for a charge equal to the accretion limit $\bigl(\a Q_0^2 =
m^2 M^2\bigr)$, both the expansion parameters reduce simply to
$(Mm)^{-2}$.

The parameter $Mm$ controls the nature of the black hole.
For $Mm \gg 1$, it is essentially a classical, general relativistic
object and quantum effects such as we are discussing here are strongly
suppressed by powers of $(Mm)^{-2} ~=~
O\bigl( (\l_c/L)^2\bigr)$.
On the other hand, for $Mm \simeq 1$, quantum effects are important
and phenomena such as black hole evaporation are significant.
In particular, for small $Mm$, the Hawking temperature $T_H =
1/(8\p M) >  m$ and electrons may be produced as Hawking radiation.

Since the Reissner-Nordstr\"om solution has an electric field,
$e^+ e^-$ pair creation will occur. The rate of production, first
calculated by Schwinger[10], is suppressed (at $r\simeq 2M$) by
an exponential factor
$\exp -\bigl( Mm (Q/Q_0 ) \bigr)$.
Again, we see that for $Q\simeq Q_0$ the relevant parameter is $Mm$
and that for $Mm\gg 1$, pair creation is srongly suppressed.
For $Q\gg Q_0$, the regime in which the effect of the charge on
the gravitational field becomes significant, the black hole
would normally lose charge rapidly through pair creation.

This double expansion is typical of quantum field theory calculations
in a background field or at finite temperature, density
etc. The general form of
the perturbation expansion in the coupling constant is
$f_0(s) + \a f_1(s) + \a^2 f_2(s) + \ldots~$, where $f_i(s)$ are
functions of a dimensionless parameter $s$ defined as the ratio
of the background scale to a physical mass in the quantum theory.
Here, $s = R/m^2$ or $\a F^2/m^4$, whereas, for example,
in calculations
of the finite temperature effective action in electroweak theories
(see, e.g., ref.[11]) $s = T/m_W$. Critical behaviour, such as phase
transitions, occurs generally when this parameter $s$ is of $O(1)$.
However, in many cases, the present calculation being in this category,
the best we can do is give an expansion of the coefficient functions
$f_i(s)$ for small and/or large $s$ (for example, the low and high
temperature expansions of the effective potential). We emphasise
that the restriction here to weak curvature $\l_c \ll L$ is purely
technical, arising from our ability to calculate the effective action.
For strong curvatures of the order of the elementary particle masses,
the phenomena described here will be enhanced, becoming ordinary
$O(\a)$ quantum effects.

There is a further expansion inherent in eq.(2.1) in the number of
derivatives acting on the electromagnetic field strength. This is
related to the dependence of the photon propagation on its
wavelength $\l$ and is discussed carefully below.

At this point, we could in principle compute the photon propagator
from the effective action (2.1) and derive the modified light-cone
condition (1.2) directly. However, it is in practice much simpler to
follow the method of ref.[1] and compute the
equation of motion for the electromagnetic field corresponding to the
photon in the geometric optics approximation.

The equation of motion of the electromagnetic field,
$$
{\d \C \over \d A_\n} ~=~ 0
,\eqno(2.3)
$$
becomes
$$\eqalignno{
D_\m F^{\m\n}~&+~{1\over m^2} \biggl[~
4a \Bigl( F^{\m\n} D_\m R + R D_\m F^{\m\n} \Bigr)  \cr
&\hskip1cm +2b \Bigl( F^{\s\n} D_\m R^\m{}_\s + R^\m{}_\s D_\m F^{\s\n}
- F^{\s\m} D_\m R^\n{}_\s - R^\n{}_\s D_\m F^{\s\m} \Bigr)  \cr
&\hskip1cm+4c \Bigl( F^{\s\t} D_\m R^{\m\n}{}_{\s\t} +
R^{\m\n}{}_{\s\t} D_\m F^{\s\t} \Bigr) \cr
&\hskip1cm+2d \Bigl( D^2 D_\s F^{\s\n} - D_\m D^\n D_\s F^{\s\m} \Bigr)~
\biggr] \cr
&+~{1\over m^4} \biggl[   -8z \Bigl( F^{\s\t}F_{\s\t} D_\m F^{\m\n} +
2 F^{\m\n} F_{\s\t} D_\m F^{\s\t} \Bigr)  \cr
&\hskip1cm
-8y \Bigl( F^{\n\t} F_{\s\t} D_\m F^{\m\s}  + F^{\m\s} F_{\s\t}
D_\m F^{\n\t} + F^{\m\s} F^{\n\t} D_\m F_{\s\t} \Bigr)~\biggr]   \cr
= ~ 0 &{}
&(2.4) \cr }
$$

To study the motion of a photon, we now write
$$
F_{\m\n} ~=~ \bar F_{\m\n} ~+~ \hat f_{\m\n}
,\eqno(2.5)
$$
where $\bar F_{\m\n}$ is the background electromagnetic field satisfying
$\d\C/\d A\big|_{\bar F} = 0$, and linearise in $\hat f_{\m\n}$.
To leading order in $\a$, we can simply take $\bar F$ and the curvature
to be the solutions of the classical field equations. The ``back
reaction'' on the metric is a higher order correction to the
results we find here.
The characteristics of the photon propagation are then given
by the geometric optics approximation where we set
$$
\hat f_{\m\n} ~=~ f_{\m\n} e^{i\o}
,\eqno(2.6)
$$
where $f_{\m\n}$ is a slowly varying amplitude and $\o$ the rapidly
varying phase, with $k_\m = D_\m \o$ corresponding to the
photon momentum.

The electromagnetic Bianchi identity
$$
D_\l F_{\m\n} ~+~ D_\m F_{\n\l} ~+~ D_\n F_{\l\m} ~=~ 0
\eqno(2.7)
$$
applied to eq.(2.5) implies
$$
k_\l f_{\m\n} ~+~ k_\m f_{\n\l} ~+~ k_\n f_{\l\m} ~=~ 0
\eqno(2.8)
$$
and so
$$
f_{\m\n} ~=~ k_\m a_\n - k_\n a_\m
.\eqno(2.9)
$$
The field strength $f_{\m\n}$ therefore has just three independent
components. The vector $a_\m$ gives the direction of polarisation
of the photon.

The ``photon equation of motion'' is now found by substituting
(2.5) and (2.6) into eq.(2.4). Imposing several simplifying
approximations, we find
$$\eqalignno{
k_\m f^{\m\n} ~&+~ {1\over m^2} \biggl[2b R^\m{}_\s k_\m f^{\s\n}
+ 4c R^{\m\n}{}_{\s\t} k_\m f^{\s\t} \biggr]  \cr
&+ {1\over m^4} \biggl[ -16z \bar F^{\m\n} \bar F_{\s\t} k_\m f^{\s\t}
- 8y \Bigl( \bar F^{\m\s} \bar F_{\s\t} k_\m f^{\n\t}
+ \bar F^{\m\s} \bar F^{\n\t} k_\m f_{\s\t} \Bigr)~\biggr]  \cr
=~0 &{}
.&(2.10) \cr }
$$

The dominant quantum corrections to the photon propagation are
of $O\bigl(\a (\l_c/L )^2 \bigr)$ from the curvature terms
and $O\bigl(\a (\a \bar F^2/m^4 ) \bigr)$ from the
electromagnetic terms. The approximations are:

\noindent (i)~~Assuming the background gravitational and electromagnetic
fields vary with the typical curvature scale $L$, terms involving
$D_\l R_{\m\n}$, $D_\l \bar F_{\m\n}$ etc. are suppressed relative
to the leading correction by $O\bigl(\l/L\bigr)$, where $\l$
is the photon wavelength.

\noindent (ii)~~Terms with derivatives acting on $f_{\m\n}$ are
suppressed by $O\bigl(\l/L_A\bigr)$, where $L_A$ is the scale of
variation of the ``photon'' amplitude. This is the standard
leading-order geometric optics approximation.

\noindent (iii)~~Terms involving $D_\m F^{\m\n}$ within the square
brackets in eq.(2.4) simply give $O(\a)$ corrections to the
$k_\m f^{\m\n}$ term in eq.(2.10) and can be omitted provided
QED perturbation theory is reliable.

The final approximation we have made is to neglect the series of
terms in the effective action involving successively higher
powers of derivatives compensated by inverse powers of mass, i.e.,
symbolically, $\sum_p \bigl(R/m^2\bigr) \bigl(D^p/m^p\bigr) F^2$ terms.
This is valid provided the photon wavelength is large compared
to the electron Compton wavelength, i.e. $\l_c\ll \l$.
As we discuss later, this is a restriction we would like to
circumvent and indeed recent developments[12] in the heat kernel
technique for calculating the effective action may make this
possible.

For the moment, however, combining these approximations we see
that our results will be valid provided $\l_c\ll \l\ll L$.
Since the dominant quantum effect is
$O\bigl(\a (\l_c/L )^2\bigr)$, it is necessarily small.

The geometric optics approximation (ii) also deserves further
scrutiny since, as noted by Dolgov and Khriplovich[5],
the classical motion of a localised wave packet is subject to
tidal forces producing $O(\l/L)^2$ corrections to the phase and
group velocities which could dominate the quantum effect on the photon
velocity calculated here. However, to decide whether this is
relevant to the full quantum theory would require recasting our
discussion in terms of the curved space photon propagator and we
leave this for further work.

To solve eq.(2.10), we use the following general method[1].
Multiplying by $k^\l$, antisymmetrising on $\l$ and $\n$, and using
the Bianchi identity gives
$$
k^2 f^{\l\n} ~-~ 2 k^{[\l}  \{\ldots \}^{\n ]} ~=~ 0
,\eqno(2.11)
$$
where $\{\ldots \}$ denotes the square bracket terms
in eq.(2.10). These are a set of linear equations for the components
of $f^{\l\n}$, with coefficients of $O\bigl(k^2\bigr)$.

Now introduce a local Lorentz frame via the vierbeins $e^a{}_\m$,
$a=0,1,2,3$, so that
$$
g_{\m\n}~=~ \h_{ab} e^a{}_\m e^b{}_\n
,\eqno(2.12)
$$
where $\h_{ab} = {\rm diag}(-1,1,1,1)$. In this frame, eq.(2.11)
becomes a set of linear equations for the field strength components
$f^{ab} = k^a a^b - k^b a^a$, where the directions of $a^a$ determine
the polarisation.

It is useful to introduce the following notation for the antisymmetric
combination of vierbeins,
$$
U^{ab}_{\m\n} ~=~ e^a{}_\m e^b{}_\n - e^b{}_\m e^a{}_\n
.\eqno(2.13)
$$
Contracting the field tensor $f^{\l\n}$ with $U^{ab}_{\m\n}$
projects the Lorentz components, i.e.
$f^{ab}  =  (1/2) f^{\m\n} U^{ab}_{\m\n}$.
The appropriate choice of a linearly independent set of three
components will depend on the specific background considered.
For example, contracting eq.(2.11) with $U^{01}_{\m\n}$,
$U^{02}_{\m\n}$ and $U^{23}_{\m\n}$ provides a set of three
linearly independent equations for $f^{01}$, $f^{02}$ and $f^{23}$.
Writing these equations in matrix form, the eigenvalues
give the light-cone conditions (1.2) on the photon momentum
while the eigenvectors determine the polarisations.

In the following section we carry through this programme for the
Reissner-Nordstr\"om metric characterising a charged black hole.
Appendix A contains the analogous discussion for a general
anisotropic electromagnetic background.

\vfill\eject

\noindent{\bf 3.~Photon Propagation in Reissner-Nordstr\"om Spacetime}
\vskip0.5cm
The Reissner-Nordstr\"om metric is an exact classical solution of
the coupled Einstein-Maxwell equations. The exterior region beyond
the horizon describes the gravitational and electromagnetic fields
of a charged black hole.

The Einstein-Maxwell equations are
$$
R_{\m\n} ~-~ {1\over2} R g_{\m\n} ~=~ 8\p T_{\m\n}
,\eqno(3.1)
$$
where the electromagnetic energy-momentum tensor is
$$
T_{\m\n} ~=~ F_{\m\s} F^\s{}_\n - {1\over4} g_{\m\n} F_{\s\t} F^{\s\t}
,\eqno(3.2)
$$
and
$$
D_\m F^{\m\n} ~=~ 0
.\eqno(3.3)
$$

The Reissner-Nordstr\"om metric in standard Schwarzschild coordinates
is[13]
$$
ds^2 ~=~ -{V(r)}^2 dt^2 +{V(r)}^{-2} dr^2 + r^2\bigl(d\o^2 +
\sin^2\o d\phi^2 \bigr)
,\eqno(3.4)
$$
where
$$
V(r) = 1 - {2M\over r} + {Q^2\over 4\p r^2}
.\eqno(3.5)
$$
There is an event horizon at
$r_+ = M\Bigl(1+\sqrt{1-{Q^2\over 4\p M^2}}\Bigr)$.
For a charge $Q\simeq Q_0$, $r_+$ is extremely close to the
Schwarzschild radius $2M$.

A local Lorentz frame can be defined with vierbeins
$$
e^a{}_\m ~=~ {\rm diag}\bigl(V(r),{V(r)}^{-1}, r, r\sin\o\bigr)
\eqno(3.6)
$$
and a basis of 1-forms
$e^a = e^a{}_\m dx^\m$
orthonormal w.r.t.~$\h_{ab}$.
The connection is found from the torsion-free Cartan equation
$$
de^a + \w^a{}_b \wedge e^b ~=~ 0
\eqno(3.7)
$$
and the curvature 2-form
$R^a{}_b   \equiv  {1\over2} R^a{}_{bcd} e^c \wedge e^d$
from
$$
R^a{}_b ~=~d\w^a{}_b + \w^a{}_c \wedge \w^c{}_d
.\eqno(3.8)
$$

The non-zero components of the curvature tensor are
$$\eqalignno{
R^0{}_{101} ~&=~ -\bigl(V V''+ {V'}^2 \bigr) \cr
R^0{}_{202} ~&=~ R^0{}_{303} ~=~R^1{}_{212} ~=~ R^1{}_{313}
{}~=~ - {1\over r} V V' \cr
R^2{}_{323} ~&=~ {1\over r^3} \bigl(1 - V^2\bigr)
&(3.9) \cr }
$$
plus other components related to these by standard symmetries.
The Riemann curvature can therefore be expressed as
$$\eqalignno{
R_{\m\n\s\t} ~=~ &- (A-B) \bigl(g_{\m\s} g_{\n\t} -
g_{\m\t} g_{\n\s}\bigr) \cr
&- (3A-4B) U^{01}_{\m\n} U^{01}_{\s\t}
{}~+~ (3A-2B) U^{23}_{\m\n} U^{23}_{\s\t}
,&(3.10) \cr }
$$
where
$$
A = {M\over r^3}, \hskip2cm
B = {1\over 4\p} {Q^2\over r^4}
.\eqno(3.11)
$$

The electromagnetic field strength is
$$
F_{\m\n} ~=~ - {1\over 4\p} {Q\over r^2} U^{01}_{\m\n}
,\eqno(3.12)
$$
describing a radial electric field.

Now consider the equation of motion for photon propagation.
To simplify notation, define
$$
\ell_\n = k^\m U^{01}_{\m\n}, \hskip2cm
n_\n = k^\m U^{02}_{\m\n}, \hskip2cm
m_\n = k^\m U^{23}_{\m\n}
.\eqno(3.13)
$$
Now contract eq.(2.11) successively with $U^{01}_{\l\n}$,
$U^{02}_{\l\n}$ and $U^{23}_{\l\n}$~ (or, equivalently, contract
eq.(2.10) with $\ell_\n$, $n_\n$ and $m_\n$).
This gives the following system of linear
equations\footnote{$^*$}{\eightpoint
In fact, the $k^2$ terms in the diagonal entries are all
multiplied by a factor $\Bigl[1 + {2bB\over m^2}
- {8c(A-B)\over m^2} \Bigr]$, but this merely gives a higher order
correction of $O(\a^2)$ in the light-cone condition
and must be dropped for consistency.

The following formulae, derived using the Bianchi identity, are
useful in simplifying these equations:
$$\eqalignno{
k_\m \ell_\n f^{\m\n} &= k^2 f^{01}  \hskip2cm
k_\m n_\n f^{\m\n} = k^2 f^{02} \hskip2cm
k_\m m_\n f^{\m\n} = k^2 f^{23} \cr
\ell_\n \ell^\t U^{01}_{\t\s} f^{\s\n} &= -\ell^2 f^{01} \hskip1.4cm
n_\n \ell^\t U^{01}_{\t\s} f^{\s\n} = -\ell^2 f^{02} \hskip1.3cm
m_\n \ell^\t U^{01}_{\t\s} f^{\s\n} = -\ell^2 f^{23} \cr
\ell^\n \ell_\s U^{01}_{\n\t} f^{\s\t} &= \ell^2 f^{01} \hskip1.6cm
n^\n \ell_\s U^{01}_{\n\t} f^{\s\t} = \ell . n f^{01} \hskip1.4cm
m^\n \ell_\s U^{01}_{\n\t} f^{\s\t} = \ell . m f^{01}
&{} \cr }
$$  },
$$
\left(\matrix{k^2+(\a+\b+\d+2\e)\ell^2 &0 &\c \ell . m \cr
(\b+\d+\e)\ell . n &k^2 +(\a+\e)\ell^2 &\c m . n \cr
(\b+\d+\e)\ell . m &0 &k^2 + (\a+\e)\ell^2 +\c m^2 \cr}\right)
\left(\matrix{f^{01} \cr f^{02} \cr f^{23} \cr }\right)
{}~=~0
,\eqno(3.14)
$$
where
$$\eqalignno{
\a &= {4b\over m^2}B \hskip2cm
\b = -{8c\over m^2} (3A-4B) \hskip2cm
\c = {8c\over m^2} (3A-2B) \cr
\d &= -{32 z\over m^4}{1\over 4\p} B \hskip1.6cm
\e = -{8y\over m^4}{1\over4\p} B
.&(3.15) \cr  }
$$
In terms of Lorentz frame components,
$\ell^2 = k^0k^0-k^1k^1$,~ $n^2 = k^0k^0-k^2k^2$,~
$m^2 = k^2k^2+k^3k^3$~ and
$\ell . m = 0$,~ $\ell . n = - k^1 k^2$,~ $m . n = - k^0 k^3$.
Notice that $\ell . m = 0$ in eq.(3.14).

Setting the determinant to zero in eq.(3.14) gives the
condition
$$
\bigl(k^2 + (\a+\b+\d+2\e)\ell^2\bigr)~
\bigl(k^2 + (\a+\e)\ell^2 + \c m^2\bigr)~
\bigl(k^2 + (\a+\e)\ell^2\bigr) ~=~ 0
.\eqno(3.16)
$$
So we have in general three roots, viz.
\vskip0.2cm
\line{\noindent (i)~~$k^2 + (\a+\b+\d+2\e)\ell^2 = 0$, \hfill}
\line{\noindent corresponding to
$f_{ab  } \propto \bigl(k_a  \ell_b  - k_b  \ell_a  \bigr)$ \hfill}
\vskip0.2cm
\line{\noindent(ii)~~$k^2 + (\a+\e)\ell^2 + \c m^2 = 0$, \hfill}
\line{\noindent corresponding to
$f_{ab  } \propto \bigl(k_a  m_b  - k_b  m_a \bigr)$ \hfill}
\vskip0.2cm
\line{\noindent(iii)~~$k^2 + (\a+\e)\ell^2 = 0$, \hfill}
\line{\noindent corresponding to
$f_{ab  } \propto \bigl(k_a  a_b  -  k_a  a_b  \bigr)$,
where $a_a  = m^2 \ell . m \ell_a
+ \ell^2 m . n m_a  - \ell^2 m^2 n_a $. \hfill}
\vskip0.5cm
These three roots are the generalised light-cone conditions
anticipated in sect.~1. To understand their implications, it is
simplest to consider two special cases:
\vskip0.5cm
\noindent (a) Radial photon motion:

\noindent Set $k^2 = k^3 = 0$. The solution (i) just produces
an overall factor multiplying $\h_{ab}k^a k^b$, i.e.
$$
\bigl(1 - (\a+\b+\d+2\e)\bigr) \bigl(-k^0 k^0 + k^1 k^1\bigr) ~=~0
.\eqno(3.17)
$$
So the light cone is unchanged and the photon velocity
$\bigl|k^0/k^1\bigr| = 1$.

The solutions (ii) and (iii) are degenerate
and also give $\bigl|k^0/k^1\bigr| = 1$.
We therefore find that for radial photon motion there is no
change in the light cone or the photon velocity for either
physical polarisation.
\vskip0.5cm
\noindent (b) Orbital photon motion:

\noindent Now set $k^1 = k^2 = 0$ and consider photon propagation
in the orbital ($\phi$) direction $k^3 \ne   0$.
\noindent (i) The first root gives the modified light cone
$$
-\bigl(1 - (\a+\b+\d+2\e)\bigr)k^0k^0 ~+~ k^3 k^3 ~=~ 0
,\eqno(3.18)
$$
i.e. a photon velocity
$$
\Bigl|{k^0\over k^3}\Bigr| ~=~ 1 + {1\over2} (\a+\b+\d+2\e)
,\eqno(3.19)
$$
corresponding to radial polarisation, i.e. $a^a \propto \d^{a1}$.
\vskip0.3cm
\noindent(ii) The second root gives a light cone
$$
-\bigl(1 - (\a+\e) \bigr) k^0 k^0 ~+~ (1+\c) k^3 k^3 ~=~ 0
,\eqno(3.20)
$$
i.e.
$$
\Bigl|{k^0\over k^3}\Bigr| ~=~ 1 + {1\over2}(\a+\c+\e)
,\eqno(3.21)
$$
corresponding to the other transverse polarisation,
$a^a \propto \d^{a2}$.
\vskip0.3cm
\noindent(iii) The third root, giving $\bigl|k^0/k^3\bigr| =
1 +{1\over2}(\a+\e)$, corresponds to the unphysical photon
polarisation $a^a \propto \bigl(k^1 \d^{a0} - k^0 \d^{a1}\bigr)$.

Substituting back using eqs.(3.15) and (2.2), we therefore find
$$
\Bigl|{k^0\over k^3}\Bigr|_{r~{\rm pol}}  ~=~
1 + {1\over30} {\a\over \p} {A\over m^2}
+ {1\over36} {\a\over\p} {B\over m^2}
- {2\over45} {\a\over\p} {\a B\over m^4}
\eqno(3.22)
$$
and
$$
\Bigl|{k^0\over k^3}\Bigr|_{\o~{\rm pol}} ~=~
1 - {1\over30} {\a\over \p} {A\over m^2}
+{13\over180}{\a\over\p} {B\over m^2}
- {7\over90} {\a\over\p} {\a B\over m^4}
.\eqno(3.23)
$$
These are the results quoted in the introduction, eq.(1.1).
The three correction terms correspond respectively to the
gravitational contribution identical to the Schwarzschild case,
the indirect effect of the charge via its induced gravitational
field and the direct electromagnetic contribution.
The latter is always negative -- the electromagnetic field reduces
the velocity of both photon polarisations (see Appx.~A).
The induced gravitational contribution in this case is positive
for both polarisations. However, as described earlier, this
is hugely suppressed by an inverse power of $\a/m^2$ relative to
the direct electromagnetic effect.

The remarkable feature of eqs.(3.22) and (3.23) is the opposite
sign of the gravitational contribution for the two different
polarisation states. This was the result discovered in ref.[1].
It can be traced back to the anisotropy in the curvature tensor,
in particular the relative sign of the $U^{01}_{\m\n}U^{01}_{\s\t}$
and $U^{23}_{\m\n}U^{23}_{\s\t}$ terms in eq.(3.10).

The question of whether photon propagation is or is not
superluminal in the Reissner-Nordstr\"om metric therefore depends
on the relative magnitude of the corrections ${1\over30}{\a\over\p}
{A\over m^2}$ and $-{2\over45}{\a\over\p}{\a B\over  m^4}$
in eq.(3.22). This is best phrased in the parametrisation of
eq.(1.1) -- superluminal propagation requires
$$
{r\over 2M} ~>~ {2\over3} \Bigl({Q\over Q_0}\Bigr)^2
,\eqno(3.27)
$$
which is satisfied everywhere outside the horizon for
charges less than the accretion limit, $Q<Q_0$.

\vfill\eject

\noindent{\bf 4. Discussion}
\vskip0.5cm
In this paper, we have shown how photon propagation is modified
in the Reissner-Nordstr\"om spacetime by vacuum polarisation
effects depending on the anisotropy of the background gravitational
and electromagnetic fields. The gravitational field may either
increase or decrease the photon velocity from $c$ depending
on its direction and polarisation, whereas the electromagnetic
field always reduces the velocity. For black hole charges
approximately equal to the accretion limit, there is a
balance between the two corrections with superluminal propagation
possible for orbital directed photons at or outside the horizon.

The fundamental questions, however, remain those confronted by
Drummond and Hathrell in their original paper[1]. The most serious is
the question of whether a spacelike photon momentum given by
the light-cone condition (1.2) necessarily involves a causal
paradox, i.e.~the possibility of a spatially closed motion
backwards in time. As emphasised in ref.[1], {\it two} conditions
have to be met to establish a violation of causality in the usual context
of special relativity, viz.~the existence of signal propagation
with spacelike momentum {\it and} Lorentz invariance. Briefly, the
argument is as follows. Suppose $A$ sends a spacelike signal to $B$.
To establish a causal paradox, $B$ must be able to return a signal
to the past world line of $A$, which requires a signal backwards in
time in this frame. Now, since the first signal $A\rightarrow B$
is spacelike, it is possible to find a new frame in which it is
indeed backwards in time. However, in order to conclude from this that
a signal $B\rightarrow A$ can be sent which is backwards in time
in the original frame, we must assume Lorentz {\it invariance}.
This final assumption is not valid in our context, since
eq.(1.2) determining the nature of the signal
propagation is a frame-dependent, Lorentz non-invariant condition.

Another crucial issue is whether the phenomenon of ``faster than light''
propagation is observable, even in principle. The essential point
here[1] is that since the maximum time     available in the
curved spacetime over which to measure a signal is ordinarily of
$O(L)$, any length discrepancy due to the corrections to the
photon velocity is roughly of order $\d s \sim L \d v \sim
\a {\l_c\over L} \l_c \ll \l_c$. But the photon wavelength $\l$
is restricted in our calculation to be much greater than $\l_c$,
so it is not at all clear how this distance $\d s$ can be resolved.
We are therefore forced to conclude that a direct measurement
of superluminal propagation may be impossible in the parameter
range $\l_c \ll \l \ll L$.

This gives added motivation to try and extend the analysis of
photon propagation to the short  wavelength regime $\l \ll \l_c$
and/or to the strong curvature regime $L \simeq\l_c$.
In terms of the effective action, relaxing the restriction
$L \gg \l_c$ would involve summing the series of higher powers in
the curvature, i.e.~terms of the form
$\sum_p  \bigl(R^p/m^{2p}\bigr) F^2$.
Unfortunately, this seems to be beyond all known techniques,
except perhaps for very special symmetric spaces.
On the other hand, relaxing the condition $\l \gg \l_c$ requires
summing terms of the type
$\sum_p \bigl(R/m^2\bigr)\bigl(D^p/m^p\bigr) F^2$.
This resummation has been discussed in a recent paper on the
heat kernel derivation of the effective action by Barvinsky
et al.[12] (see also refs.[14]). We are currently investigating
whether these results enable us to describe photon propagation
for arbitrary $\l$.

Opinions in the literature[1,4-8] are divided as to whether the
photon velocity for the superluminal polarisations would increase
or decrease as $\l$ becomes small. Based on a preliminary analysis
of ref.[12] or a consideration of the large external momentum
behaviour of the appropriate Feynman diagrams for vacuum polarisation,
we tend to the view that for small photon wavelengths, the r\^ole
of the electron mass is taken over by the photon momentum so that
the velocity correction for the parameter range $\l \ll \l_c \ll L$
is of $O\bigl(\a (\l/L)^2\bigr)$.
All polarisations would therefore travel with velocity equal to $c$
in the limit $\l \rightarrow 0$. The objection to this scenario
is that it requires the absorptive part  in the usual dispersion
relation for the refractive index to be negative for the superluminal
polarisations while remaining conventionally positive for the
subluminal polarisation states. Such admittedly unconventional
behaviour does not, however, appear to us to be ruled out
although we cannot at present give a mechanism to understand to it.
In any case, the issue should be settled by an explicit calculation.

Finally, we emphasise again what we see as ultimately the most
important aspect of this phenomenon, viz.~the violation of local
Lorentz invariance by the tidal gravitational effects arising
through vacuum polarisation in an interacting quantum field theory.
Such a fundamental result must surely have far-reaching
implications for quantum gravity.

\vskip2cm

\noindent{\bf Acknowledgements}
\vskip0.5cm
We would like to thank I.~Drummond, G.~Veneziano and Ph.~Zaugg
for interesting discussions.
One of us (G.M.S.) is especially grateful to Prof.~H.~Ruegg for
hospitality at the University of Geneva while this work was being
completed.

\vfill\eject

\noindent{\bf Appendix A.~~~Electromagnetic Birefringence}
\vskip0.5cm
We show here how to calculate the velocity of light in an arbitrary
anisotropic (but only weakly inhomogeneous) electromagnetic field
in flat spacetime. This generalises Adler's original calculation
of electromagnetic birefringence[3]. From the perspective of this
paper, it is interesting to consider this general case in order
to check that the resulting photon velocity is always less than $c$
independent of the polarisation, despite the background field
anisotropy. This confirms that the ``faster than light'' phenomenon
is a gravitational effect.

We use the same techniques described in sects.~2 and 3. The relevant
effective action is the familiar Euler-Heisenberg expression,
viz.~eq.(2.1) without the curvature terms. In the geometric optics
approximation, the ``photon equation of motion'' (cf.~eq.(2.10))
is simply
$$
k_a  f^{ab  } + {1\over2} \d F^{ab  } F_{cd  } k_a  f^{cd  }
+\e \Bigl(F^{ac  } F_{cd  } k_a  f^{bd  }
+ F^{ac  } F^{bd  } k_a  f_{cd  } \Bigr)
{}~=~0
,\eqno(A.1)
$$
where here we define
$$
\d = -{32 z\over m^4} \hskip2cm
\e = - {8 y\over m^4}
.\eqno(A.2)
$$
The background electromagnetic field is
$$
F_{cd} ~=~
\left(\matrix{0 &-E_1&-E_2&-E_3 \cr
E_1&0&B_3&-B_2 \cr
E_2&-B_3&0&B_1 \cr
E_3&B_2&-B_1&0 \cr}\right)
,\eqno(A.3)
$$
that is,
$$
F_{cd} ~=~-E_1 U^{01}_{cd}  - E_2 U^{02}_{cd} - E_3 U^{03}_{cd}
+ B_3 U^{12}_{cd} - B_2 U^{13}_{cd} + B_1 U^{23}_{cd}
,\eqno(A.4)
$$
where in analogy to eq.(2.13) we have defined
$$
U^{ab}_{cd} ~=~ \d^a_c \d^b_d - \d^b_c \d^a_d
.\eqno(A.5)
$$

{}From the Bianchi identity, we have
$$
f^{ab} ~=~ k^a a^b - k^b a^a
.\eqno(A.6)
$$
This shows that for given $k^a$, only three of the $f^{ab}$
are independent. We take these to be $f^{01}$, $f^{02}$ and
$f^{03}$ and express the others as
$$\eqalignno{
f^{23}&= {1\over k^0}\bigl(k^2 f^{03} - k^3 f^{02}\bigr)  \cr
f^{13}&= {1\over k^0}\bigl(k^1 f^{03} - k^3 f^{01}\bigr)  \cr
f^{12}&= {1\over k^0}\bigl(k^1 f^{02} - k^2 f^{01}\bigr)
.&(A.7) \cr }
$$

We also define
$$
\ell_d = k^c U^{01}_{cd} \hskip2cm
n_d = k^c U^{02}_{cd} \hskip2cm
p_d = k^c U^{03}_{cd}
.\eqno(A.8)
$$
Again, the other possibilities are not independent:
$$\eqalignno{
m_d&= k^c U^{23}_{cd} = {1\over k^0}\bigl(k^2 p_d - k^3 n_d\bigr)  \cr
q_d&= k^c U^{13}_{cd} = {1\over k^0}\bigl(k^1 p_d - k^3 \ell_d\bigr) \cr
r_d&= k^c U^{12}_{cd} = {1\over k^0}\bigl(k^1 n_d - k^2 \ell_d\bigr)
.&(A.9) \cr }
$$

We therefore find the following useful formulae:
$$
-{1\over2}F_{ab}f^{ab} ~=~ \tilde E_1 f^{01} + \tilde E_2 f^{02}
+ \tilde E_3 f^{03}
\eqno(A.10)
$$
and
$$
-F_{ab} k^a ~=~ \tilde E_1 \ell_b + \tilde E_2 n_b
+ \tilde E_3 p_b
,\eqno(A.11)
$$
where
$$
\tilde E_1~=~ E_1 - {k^3\over k^0} B_2 + {k^2\over k^0} B_3
,\eqno(A.12)
$$
etc., together with
$$
f^{ab} \ell_a  =  - k^b f^{01}  \hskip2cm
f^{ab} n_a = -k^b f^{02}  \hskip2cm
f^{ab} p_a = -k^b f^{03}
,\eqno(A.13)
$$
which follow from the Bianchi identity.

With these preliminaries, a straightforward calculation
along the lines of sect.~3 shows that the equation of motion (A.1)
for $f^{ab}$ reduces to the following matrix equation for the
independent components $f^{01}$, $f^{02}$ and $f^{03}$:
$$\eqalignno{
&\left(\matrix{k^2+(\d+\e)\tilde E_1\BOB +\e X
&(\d+\e)\tilde E_2\BOB &(\d+\e)\tilde E_3\BOB \cr
(\d+\e)\tilde E_1\BSB &k^2+(\d+\e)\tilde E_2\BSB +\e X
&(\d+\e)\tilde E_3\BSB \cr
(\d+\e)\tilde E_1\BCB &(\d+\e)\tilde E_2\BCB
&k^2 +(\d+\e)\tilde E_3\BCB +\e X \cr}\right)
\left(\matrix{f^{01}\cr f^{02}\cr f^{03} \cr}\right) \cr
&=~0
,&(A.14) \cr }
$$
where
$$\eqalignno{
\BOB ~&=~\tilde E_1 \ell^2 + \tilde E_2 \ell . n +
\tilde E_3 \ell . p \cr
\BSB ~&=~\tilde E_1 \ell . n + \tilde E_2 n^2 +
\tilde E_3    n . p \cr
\BCB ~&=~\tilde E_1 \ell . p + \tilde E_2 n . p  +
\tilde E_3 p^2
&(A.15) \cr }
$$
and
$$\eqalignno{
X~&=~\tilde E_1 \BOB + \tilde E_2 \BSB + \tilde E_3 \BCB \cr
&=~\bigl(k^0E_3 - k^2B_1 + k^1B_2\bigr)^2
+\bigl(k^0E_2 - k^1B_3 + k^3B_1\bigr)^2 \cr
&~~~+\bigl(k^0E_1 - k^3B_2 + k^2B_3\bigr)^2
-\bigl(k^1E_1 + k^2E_2 + k^3E_3\bigr)^2
.&(A.16) \cr }
$$

Setting the determinant to zero, we find the following equation
analogous to eq.(3.16):
$$
\bigl(k^2 + \e X\bigr)^2 \bigl(k^2 + (\d+2\e)X\bigr) ~=~ 0
.\eqno(A.17)
$$
There are two coincident roots, corresponding to the modified
light cone
$$
k^2 + \e X ~=~ 0
\eqno(A.18)
$$
and one other corresponding to
$$
k^2 + (\d + 2\e) X ~=~ 0
.\eqno(A.19)
$$

To analyse the light velocities, we can now with no loss of generality
choose a specific direction of photon propagation, say
$k^1 = k^2 = 0$ with  $k^0,k^3 \ne 0$.
A further short calculation now shows that the photon velocities
corresponding to the two distinct roots (A.18) and (A.19)
are
$$
\Bigl|{k^0\over k^3}\Bigr| ~=~ 1 + {1\over2} \e
\Bigl[ \bigl(E_1-B_2\bigr)^2 + \bigl(E_2+B_1\bigr)^2 \Bigr]
\eqno(A.20)
$$
and
$$
\Bigl|{k^0\over k^3}\Bigr| ~=~ 1 + {1\over2} (\d +2\e)
\Bigl[ \bigl(E_1-B_2\bigr)^2 + \bigl(E_2+B_1\bigr)^2 \Bigr]
,\eqno(A.21)
$$
the only difference being the coefficients.

The important observation is that the dependence on the background
electric and magnetic fields takes the form of a sum of two perfect
squares and is therefore always positive. Since both the coefficients
$\e$ and $\d+2\e$ from the Euler-Heisenberg effective action are
negative, we find that the photon velocity for an arbitrary anisotropic
electromagnetic background field is always less than $c$.

\vfill\eject

\noindent{\bf References}
\vskip0.5cm

\settabs\+\ [&1] &G.M.Shore  \cr

\+\ [&1] &I.T. Drummond and S.J. Hathrell, Phys. Rev. D22 (1980)
343 \cr
\+\ [&2] &P. Mende, lecture at the Workshop on String Quantum Gravity,\cr
\+\ &{}  &\hskip1cm Erice, 1992. Brown Univ.~preprint HET-875 \cr
\+\ [&3] &S.L. Adler, Ann. Phys. (N.Y.) 67 (1971) 599 \cr
\+\ [&4] &Y. Ohkuwa, Prog. Theor. Phys. 65 (1981) 1058 \cr
\+\ [&5] &A.D. Dolgov and I.B. Khriplovich, Sov. Phys. JETP 58 (1983)
671 \cr
\+\ [&6] &K. Scharnhorst, Phys. Lett. B236 (1990) 354 \cr
\+\ [&7] &G. Barton, Phys. Lett. B237 (1990) 559 \cr
\+\ [&8] &G. Barton and K. Scharnhorst, Univ.~of
Sussex preprint 9242 (1992)\cr
\+\ [&9] &G.W. Gibbons, Comm. Math. Phys. 44 (1975) 245 \cr
\+ [1&0] &J. Schwinger, Phys. Rev. 82 (1951) 664 \cr
\+ [1&1] &G.M. Shore, Ann. Phys. (N.Y.) 128 (1980) 376 \cr
\+ [1&2] &A.O. Barvinsky, Yu.V. Gusev, G.A. Vilkovisky and
V.V. Zhytnikov, \cr
\+\ &{}  &\hskip1cm Univ.~of Manitoba preprint (1993) \cr
\+ [1&3] &C.W. Misner, K.S. Thorne and J.A. Wheeler, `Gravitation', \cr
\+\ &{}  &\hskip1cm (Freeman, San Francisco) 1973 \cr
\+ [1&4] &G.A. Vilkovisky, CERN preprint TH.6392/92 \cr
\+\ &{}  &A.O. Barvinsky and G.A. Vilkovisky, Phys. Rep. 119 (1985)
1 \cr
\+\ &{}  &A.O. Barvinsky and G.A. Vilkovisky, Nucl. Phys. B282 (1987)
163, \cr
\+\ &{}  &\hskip1cm {\it ibid.} B333 (1990) 471,
{\it ibid} B333 (1990) 512 \cr

\vfill\eject

\bye